\documentclass[review,3p]{elsarticle}

\usepackage{tabulary}
\usepackage{booktabs}
\usepackage{lineno,hyperref}
\usepackage{amssymb}
\usepackage{amsmath}
\usepackage{url}
\usepackage{graphicx}
\usepackage{wrapfig}
\usepackage{lscape}
\usepackage{rotating}
\usepackage{epstopdf}

\usepackage{amsfonts}
\usepackage{url}
\usepackage{amssymb}
\usepackage[justification=centering]{caption}
\usepackage{multirow}
\usepackage{color}

\usepackage{scrextend}
\usepackage[english]{babel}
\usepackage{blindtext}
\modulolinenumbers[200]
\usepackage{adjustbox}
\usepackage{color}
\usepackage[linesnumbered,ruled,vlined]{algorithm2e}

\usepackage{graphicx}

\usepackage{subfig}
\usepackage{longtable}

\begin{document}
	\begin{frontmatter}

		\title{EDGF: Empirical dataset generation framework for wireless network networks}

		
		\author{Dinesh Kumar Sah, \hspace{0.0cm}  Praveen Kumar Donta \hspace{0.0cm} Tarachand Amgoth}
		\address{Department of Computer Science and Engineering \\
			Indian Institute of Technology (Indian School of Mines), Dhanbad\\
			Jharkhand, India-826004\\
			{dksah.iitd@gmail.com, \hspace{0.05cm} praveeniitism@gmail.com \hspace{0.05cm} tarachand.ism@gmail.com}}

		\begin{abstract}
			In wireless sensor networks (WSNs), simulation practices, system models, algorithms, and protocols have been published worldwide based on the assumption of randomness. The applied statistics used for randomness in WSNs are broad in nature, e.g., random deployment, activity tracking, packet generation, etc. Even though with adequate formal and informal information provided and pledge by authors, validation of the proposal became a challenging issue. The minuscule information alteration in implementation and validation can reflect the enormous effect on eventual results.  In this proposal, we show how the results are affected by the generalized assumption made on randomness. In sensor node deployment, ambiguity arises due to node error-value ($\epsilon$), and it's upper bound in the relative position is estimated to understand the delicacy of diminutives changes.
			Moreover, the effect of uniformity in the traffic and contribution of scheduling position of nodes also generalized. We propose an algorithm to generate the unified dataset for the general and some specific applications system models in WSNs. The results produced by our algorithm reflects the pseudo-randomness and can efficiently regenerate through seed value for validation. 
			
		\end{abstract}
		\begin{keyword}{Dataset generation , random deployment, wireless network, clustering, traffic data.}
		\end{keyword}

	\end{frontmatter}--
	\section{Introduction}
	
	In wireless sensor networks (WSN) research, the  correctness of simulation practices and results are important because of the limitation occurs implementations and testbeds.  There are many pseudo-random numbers generators (PNRG) are available such as simulators Mersenne Twister \cite{matsumoto1998mersenne}, Xorshift \cite{owen1998latin, marsaglia2003xorshift}, linear congruential generators \cite{de1988parallelization, fishman2013monte, l1990random, l1999tables, l1993search} and so
	forth \cite{l2017history,l2012random}.
	The PNRG produces 32-bit or 64-bit words, which can further decode as integer based on the requirement. In most of the simulation or programming tools, ``Mersenne Twister'' \cite{matsumoto1998mersenne} has been used to generate the random numbers (RN). The problem that arises in PRNG is that even with the same seed value, the reproduction of the same sequence is almost impossible. Several existing proposals in WSNs result, use random deployment to validate connectivity, packet generation, and much more use this function regularly. These practices raise a concern about the validation of results. Often formal description seems missing in the proposal, and provided function not enough to reproduce. Moreover, the exploration of random processes provided by simulator and programming tools is also tedious, creating hurdles for authors. 
	
	This work proposes a new dataset generation framework for WSNs referred to as empirical data generation framework (EDGF). The objective is to provide the unified datasets which are empirical for validation purpose. Our function's main contribution is that it uses a modified version of the linear congruential function with internal seed value based on the user input. We also observe some useful instances through experiments: the seed constants, provide string deployment automatically and report in the result analysis. Moreover, our proposal is  independent of the limitation imposed on the congruential generator 	\cite{james1990review}  of selection of prime $m$ and selection of $a$ as a primitive element modulo $m$.
	
	\subsection{Contributions}
	
	The contributions of the work are summarized as follows. 
	
	\begin{itemize}
		\item[1.] We propose a new deterministic algorithm to generate datasets for sensor node deployment and packet generation. The objective is to provide the dataset for 2-D deployment coordinates, and the traffic matrix corresponded to the deployed nodes. 
		
		\item[2.] Deployment data generation with the critical aspects of function and seed value. 
		
		\item[3.] Traffic generation data for uniform and exponential traffic matrix. The significance of exponential traffic is that it is needed for event-driven applications. 
		
		\item[4.] Randomness property validation of EDGF data with KS-test, $\chi^2$-test and auto-correlation test to assure the uniformity and randomness.
		
		\item[5.] Illustration of datasets produced by EDGF and highlighting important issues.
		
	\end{itemize}

	\section{Effect of random datasets in WSNs}
	
	In this section, we will cover the significance of the randomness and its effect on the validation. Here, randomness on deployment and data traffic are considered because these are standard requirements of almost all the WSNs models. However, we are in the initial stages of EDGF; therefore, we are discussing datasets for deployment and packet generation only in this work. We also observe that our framework's true potential is limitless, and many other datasets can also be generated using the EDGF for various applications in WSNs. 
	
	Most of the time, linear congruential generator and middle square method used to generate RN for deployment. In computing and programming, the algorithm can produce an extensive RN sequence and can determine by the initial input is called seed-based PRNG. BY knowing the seed value, the entire RN sequence can regenerate, which is often a requirement of computer protocols. 
	This type of RN generator is often called a pseudo-random number generator (PRNG). Moreover, the RN generated by the PRNG takes the system clock as seeds value. In our example, we are taking  2-D random values on the X-Y plane generated by Python's function (NetworkX-package) to depict the points as node' location in deployment. To proceed with the effect, we perform a K-mean clustering algorithm to form a cluster on those points. In many of the WSNs application practices, the clustering operation is often used due to its usefulness in energy, lifetime, congestion, and many more. We further estimate the euclidean distance among the cluster head and sink. After estimation, it is easy to observe the values often show the significant difference among them. Note that the distance among nodes is directly proportional to the energy consumption; transmission range estimation affects the network performance. 
	
	\begin{figure}
		\centering
		\includegraphics[width=0.5\textheight]{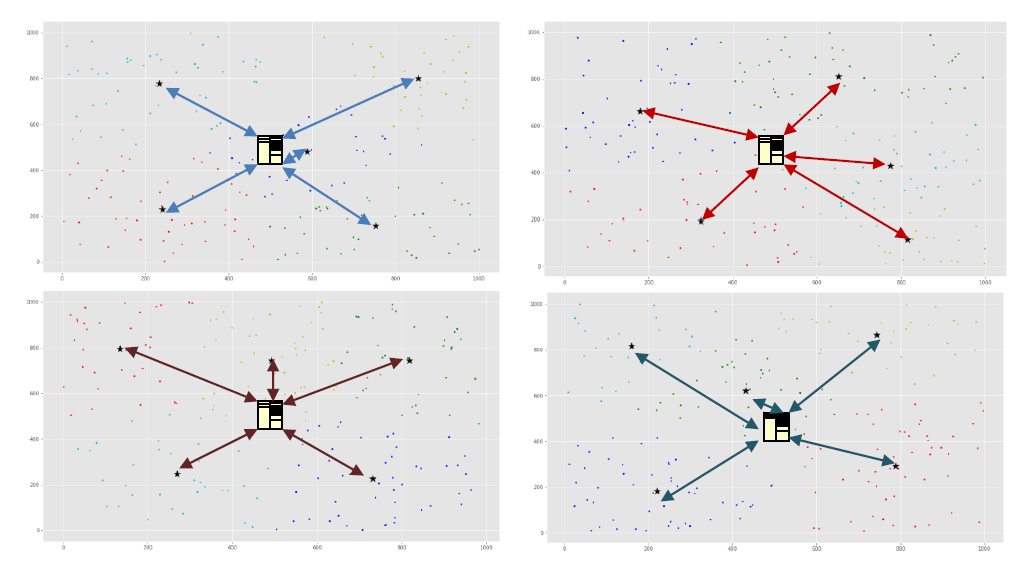}
		\caption{Cluster generated through K-mean with 4 random instances}
		\label{fig_cluster-1}
	\end{figure}
	
	To analyze the effect of randomness in data traffic generation, we consider CSMA and TDMA-MAC to explore the effects. In \cite{tay2004collision}, CSMA/$p^*$ has been considered based on the optimal probability distribution of the traffic produced by the nodes. The constraint is that the number of nodes and channel access rate of nodes is supposedly known for this protocol to enhance channel utilization enhancement. Moreover, an extension of \cite{tay2004collision} another proposal name ``sift'' \cite{jamieson2006sift} been given based on CSMA/$p^*$ which also provide better utilization in the case of availability of data at the nodes. If the data arrival of nodes is random, it is almost impossible to sense the transmission of more than two hops. The performance also starts degrading only because of the randomness.

	In Z-MAC \cite{rhee2008z}, distributed algorithm DRAND \cite{rhee2009drand} extension of centralized protocol RAND \cite{ramanathan1999unified}  used to assign the time slots to each node for synchronization to perform the node schedule further. As synchronization is one of the important aspects of node scheduling, it's highly anticipated that most of the synchronization algorithms consider the hop localization to assign the slot number. If the node position at the time of deployment is taken randomly, then the slots will vary throughout the validation. Further, to enhance the usability of unused slots, an interesting mechanism has been given. Let suppose, we have 4 node {a,b,..,j}, need to schedule in order of {a,b,c...,j}. Now all the node is the owner of its time slot and free to transmit its packet through TDMA. If any owner node is not using its slot, then those slots will be eligible for CSMA data transmission. The effect of uniform random packet generation is that suppose if any $2^{nd}$ slot and $9^{th}$ slot is not used by its owner. Then in these instance, the congestion on $2^{nd}$ will be $\sum_{leftover} packets(2 \rightarrow onwards)$ and it must be higher than $\sum_{leftover} packets(9\rightarrow onwards)$ in uniform traffic scenario. These instances can have a ripple effect in both ways, but one thing for sure that the packet transmission rate will be different in both instances. Now, at the time of validation due to congestion, the same algorithm's packet delivery rate will be different for the same random function.
	This section has shown the observable evidence of how the random function can cause a severe difference in result validation, even though using the same procedure for deployment and packet generation.

	\section{Empirical dataset generation framework }
	In this section, we develop EDGF for WSNs scenario for node deployment and packet generation. The significance of randomness in the dataset such as periods, uniformity, and independence, are covered carefully for both instances. In deployment, apart from randomness, the other issues such as localization, connectivity, outliers handling mechanism, and coverage are considered to make it as generalized as possible. To determine the packet generation of the individual nodes, the proportionality of the packet generation needs to identify first. In \cite{sah2018parametric}, the interval of the packet generation rate and its distribution throughout the networks has been shown.

	\subsection{Topological significance of datasets}
	
	In WSNs, the deployment of nodes represented as graph $G$ with nodes, $n_1,n_2,..n_i$  and the coordinate value in 2-D plane represented as $n_{1,x,y},n_{2,x,y},...,n_{i,x,y}$. The topological significance of the network follows the same properties as the graph. The graph is said to be connected if there exists a path between every pair of nodes. For a connected graph with $i$ number of nodes, $i(i-1)/2$ edges are needed to make the path between every pair of nodes. In WSNs,  very little certainty that each node falls into the range of every other node. Though the network connectivity is a different problem, many random generated functions do not care to provide the points that can form the connected graph within the sensing range $R$. One simple solution is to increase the sensing range of each node, but energy consumption will vary at the validation time. The EDGF identified these issues and pointed out the number of isolated nodes with individual sensing range values. However, EDGF is able to provide the coordinate based on a non-hierarchical manner suitable for mesh topology.
	
	We observe some critical issues like connectivity, coverage, and localization are not full-filled by the value generated through the random function because of the user's lesser control. Therefore, we are merely highlighting the issues and leaving as open problems. The nodes' output and position are provided, which can further incorporate as input to handle connectivity, coverage maximization, and localization in networks. In \cite{l1990random}, many random generators have been explored, out of which we have opted $x_n := (a(x_n-1)+c) mod m$, though changing the constant dose not going to change the vector. With this selection, only two possibilities left as either change the constant with some good suggestion given in \cite{james1990review} or combined it with some other generator integrated multiple-recursive generator (CMRG) as in \cite{l2002object}.
	Moreover, another combination is the  $a$ and $c$ value can be a real number. The possibilities  $a$ and $c$ are a real number is limitless, and floating operation is also expensive. Moreover, with some limited number of test instance and universal mathematical constant values \cite{finch2003mathematical},  the results are impressive. Though the vast exploration and validation require to check the limit of the function, it seems to pass in the initial tests performed for deployment. Our function's advantage is that it has less memory requirement as it does not need to store the constant value in term of $2^{32}$ or more to maintain the period of randomness. Moreover, our deployment's complexity is also $O(n)$, which is enough to proceed the path forward.

	In algorithm.~\ref{algo_1} to ~\ref{algo_4}, the input set is define as $n$ number of sensor nodes, $m$ area of deployment in $unit^2$ as $m \times n$ where $m$ and $n $ can be width and length, $X[0]$ is seed value for $X$ and $Y$ -coordinate, $mcV$ is the list of mathematical constant values defined internally based on the seed value given as input. Moreover, $a$ and $c$ are constant generated using the function $a = mcV[ X[0]\%|mcV| ] $ , $c = mcV\left[\left(X[0]+\frac{|mcV|}{2}\right)\%|mcV|\right]$ and   $a \neq c $. 
	Moreover, for packet generation, the number of time slots needed and defied as $t$, with the number of packets generation range, vary between the $P_1$ and $P_2$.

	\subsection{Grid deployment}
	Here, the objective is to generate the empirical dataset for two-dimensional space, such that it obey the simple rules of randomness and networks. The first requirement is it should be surely random within the range of the area of interest. The WSNs properties such as connectivity and coverage of the network should be maintained. In some instances, outliers exist in the network, so there must be the proper outliers handling mechanisms. Moreover, to deals with the outliers, the relaxation in the term of error-value $(\epsilon-value)$ 
	is being made to accommodate the outliers nodes in the network to maintain the connectivity. Meanwhile, the following relaxation is being optional in the algorithm with the concern that some instance such as mobile sink able to handle the outliers and hidden nodes. Moreover, through the experimental observations, the uniformity of node deployment is enhanced with the adoption of grid deployment. In our work, grid is define as the area such as if we plot the axis by taking the coordinate $(m/2,n/2)$. The entire grid area $G$ further divided into four sub-grids $G={g_1,g_2,g_3,g_4}$. The advantage of this assumption that it increase the connectivity and coverage of network without any separate algorithm.

	\begin{algorithm}[t]
		\DontPrintSemicolon
		{\scriptsize \KwData{(n, m, X[0])}
			\KwResult{Deployment coordinates (X,Y) in range R}
		}
		\caption{Deployment data generation without grid Algorithm
			\label{algo_1}}
		\Begin{	
			// $a$ and $c$ are constant $a \neq c $\;\;
			$a = mcV[ X[0]\%|mcV| ] $ \;
			$c = mcV\left[\left(X[0]+\frac{|mcV|}{2}\right)\%|mcV|\right]$ \;
			Y[0]=X[0] \hspace{1cm}     										//  $y$ coordinate seed value \;\;
			
			// Deployment of Sensors \;
			\For{i=1 to n}{
				X[i]=(a*X[i-1]+c)\%m \;
				Y[i]=(a*Y[i-1]+a)\%m \;
			}\;
			
			// Make Graph \;
			\For{i=1 to n}{
				G.add\_node(Point(X[i],Y[i])) \;
			}
		}			
	\end{algorithm}
	
	\begin{algorithm}[!t]
		\DontPrintSemicolon
		{\scriptsize \KwData{(n, m, X[0])}
			\KwResult{Deployment coordinates (X,Y) in an area $m \times n$}
		}
		\caption{Grid-based deployment data generation Algorithm
			\label{algo_2}}
		\Begin{	
			$a = mcV[ X[0]\%|mcV| ] $ \;
			$c = mcV\left[\left(X[0]+\frac{|mcV|}{2}\right)\%|mcV|\right]$ \;
			Y[0]=X[0] \hspace{1cm}     										//  $y$ coordinate seed value \;
			
			m1 = $\frac{m}{2}$\;
			n1 = $\frac{n}{4}$ 							//Here area dividing in to 4 grids.\;\;

			// Deployment of Sensors in Grid-1\;
			\For{i=1 to n1}{
				X[i]=(a*X[i-1]+c)\%m1 \;
				Y[i]=(a*Y[i-1]+a)\%m1 \;
			}\;
			
			// Deployment of Sensors in Grid-2\;
			\For{j=1 to n1}{
				X[i+j]= X[j] + m1 \;
				Y[i+j]= Y[j] + m1 \;
			}\;
			
			// Deployment of Sensors in Grid-3\;
			\For{k=1 to n1}{
				X[i+j+k]= X[k] + m1 \;
				Y[i+j+k]= Y[k]  \;
			}\;
			
			// Deployment of Sensors in Grid-4\;
			\For{l=1 to n1}{
				X[i+j+k+l]= X[l]  \;
				Y[i+j+k+l]= Y[l] + m1 \;
			}\;
			
			// Make Graph \;
			\For{i=1 to n}{
				G.add\_node(Point(X[i],Y[i])) \;
			}
		}			
	\end{algorithm}
	

	\section{Packet generation}
	There are two class of data or packet traffic for WSN applications, continuous monitoring  or event driven monitoring. The requirement of packet traffic may vary along with the applications.
	\subsection{Uniform Packet generation }
	
	In continuous monitoring applications, the senors send the packets in every time interval. These activity can be model as uniform distribution of packets throughout the network within the specific range.

	\subsubsection{lemma}
	
	Consider a continuous monitoring system, the packet rate of sensors model as uniform distribution traffic.

	\subsubsection{proof}
	If $x_i$ is a value taken from the TUD, then the value $a + (b-a)x_i$ follows the TUD constrain by  a and b. 
	The probability, that a TUD random variable can categorized in the interval of with the fixed range of variables is independent of the position in the interval itself ( At the same it depend upon  the interval size). To observe this, if $X$ is uniformly distribute over $ U(a,b)$ and $[x, x+d]$ is a subinterval of $[a,b]$ with fixed $d > 0,$ then $P(X \in [x,x+d])= \int_{x}^{x+d} \frac{dy}{b-a}=\frac{d}{b-a}$  which is independent of x.

	
	\begin{algorithm}
		\DontPrintSemicolon
		{\scriptsize \KwData{(n, t, $P_1$, $P_2$)}
			\KwResult{Uniform packet generation algorithm}
		}
		\caption{Uniform packet generation $r_{1,t},r_{2,t},...r_{n,t}$ for node $1,2..,n$ and for time $t$
			\label{algo_3}}
		\Begin{	
			// $P_1$ = Minimum size of packet \;
			// $P_2$ = Maximum Size of packet \;
			// $a$ and $c$ are constant $a \neq c$ \;\;
			
			X[0] = mcV[$P_2$\%$|$mcv$|$] \;
			$a = mcV[ X[0]\%|mcV| ] $ \;
			$c = mcV\left[\left(X[0]+\frac{|mcV|}{2}\right)\%|mcV|\right]$ \;\;
			\For{i=1 to n}{
				\For{j=1 to t} {
					$X[j\%t]=[a(a*X[j-1]+c)\%(P_2-P_1)]+P_1$
					$Y[i][j]=[\frac{-1}{\lambda}log(1-\frac{X[j\%t]}{P_2})\%(P_2-P_1)]+P_1$
				}
			}
			
		}			
	\end{algorithm}

	\subsection{Exponential data generation for event driven monitoring (EDM) }
	
	In EDM, the sensors is being activate only when some activity occurs in sensing range. These are the scenario when sensors can exploit the correlation to improve the network efficiency by adopting the data acquisition mechanism. In \cite{das2015correlation}, the network management has been suggested based on the spatial-temporal (ST) relation of sensors. The advantage of  adoption of ST is that it can ensure the effect of any event driven activity and  can distribute that among the nodes  exponentially. Let assume that at any time instance $t$ some event $e_t$ occurs in grid $g_i$ and in  locality $l_{2,{g_i}}$ which effect is monitored by sensor in those locality. Now, as the point of activity and individual sensor $s_i$ distance being increase, the effect start fading up. There is the possibilities that the corner most sensors of other grid  in different locality such as $l_{1,{g_{i+1}\%4}},l_{2,{g_{i+2}\%4}},l_{3,{g_{i+3}\%4}},l_{4,{g_{i+3}\%4}}$ will not even able to sense the event due to the limitation on the sensing range.  In this instance the uniform generation of data packets in simulation practices might lead to ambiguous result. Therefore, we are giving the algorithm to generate the data packet with exponential distribution.  Let assume the $F(x)$ be the exponential function distribute over random variable $x$ and rate $\lambda$ and define as:

	\begin{equation}
	\label{step_1_eq1}
	F(x)=\left\{\begin{matrix}
	1-e^{- \lambda x}, x\geq0 & \\ 
	0, x<0 & 
	\end{matrix}\right.
	\end{equation}
	Here $\lambda > 0$ is the rate parameter of the exponential.
	Now we already generated $r_{1,t},r_{2,t},...r_{n,t}$ through algorithm ~\ref{algo_3} which individual input value $r_i=1-e^{- \lambda x_i}$ for $\lambda=1$. Further, $	- \lambda x_i =ln(1-r_i)$ which will generate the individual $x_i =- \frac{1}{\lambda}ln(1-r_i)$.  (Generally $x_i$ w.r.t $r_i$). Now, the minimum effect or null effect for the nodes which are further away from the event can also prove to be random (see Appendices A.1).

	\subsubsection{lemma}
	Consider a correlation aware monitoring system, the packet rate of sensors model to the considerably far nodes is also exponential distribution traffic.

	\subsubsection{proof}
	In  Appendices A.1, the proof of minimum $\lambda$ is given which is similar to the nodes which is far from the activity point. The  $\lambda$ will be small but still the data generated w.r.t the point shows exponential distribution.

	\begin{algorithm}
		\DontPrintSemicolon
		{\scriptsize \KwData{(n, t, $P_1$, $P_2$)}
			\KwResult{Exponential packet generation (Sample result available in Appendices A.4, Table~.\ref{table_2} ) }
		}
		\caption{Exponential Packet Generation Algorithm\label{algo_4}}
		\Begin{
			// $a$ and $c$ are constant $a \neq c$ \;\;
			
			X[0][0] = mcV[$P_2$\%$|$mcv$|$] \;
			a=mcV[$P_1$\%$|$mcv$|$] \;
			$c = mcV\left[\left(X[0]+\frac{|mcV|}{2}\right)\%|mcV|\right]$ \;\;
			\For{i=1 to n}{
				\For{j=1 to t} {
					$X[i$\%$t][j$\%$t]=(a*X[(i-1)$\%$t][(j-1)$\%$t]+c)$\%$(P_2-P_1)+P_1$
				}
			}
			
		}			
	\end{algorithm}

	\section{Analysis and results}
	To establish the degree of agreement among the distribution of a sample of generated RN and theoretical uniform distribution (TUD), we are performing two well know for uniformity test and one for independence test.
	\begin{itemize}
		\item[1.] Kolmogorov-Smirnov test (KS-test) (uniformity test)
		\item[2.] $\chi^2$  test ($\chi^2$Test) (uniformity test)
		\item [3.] Autocorrelation test (independence test)
	\end{itemize}
	
	To analyize the RN generated for the test, two hypotheses made as  one support  the RN generator is indeed uniformly distributed. Then $H_0$,  can define as a null hypothesis. The other one can support  the RN generator is not uniformly distributed can represent as $H_1$, and act as alternative hypothesis. Depending upon the null hypothesis of no major difference among sample distribution and TUD we will conclude that whether the number generated through function is truly random with significance levels($\alpha$). In the experimental results, with respect to different constant $a,c$ and $KS-test(\alpha=0.01), \chi^2-test(\alpha=0.001), autocorelation(\alpha=0.01) $, the test results are tabulated in  Appendices A.3, Table~.\ref{table_1}. 
	
	In analysis, the hypothesis based test predict that whether the  two mutually exclusive statements regarding the sample to figure out which statement is more supportive to  sample data drawn from the population. 
	Further, the test result is significance only if the sample data drawn from the population is large enough or frequent enough comparative to the $H_0$ to reject the $H_0$ for overall population. There is some instance of special concern in  hypothesis test such as the assumption of $H_0=true$, limit of $\alpha$-value, nature of sample data (large value drawn from critical region). We have to understand here that,  test measurement based on $\alpha$ dose not support the acceptance as $100\%$ accuracy but to not reject the nature (randomness in our case) with individual $\alpha$. The cases where either the $\alpha$ value is change or sample change to make sure the confidence of data but it will never be $100\%$ accurate.      
	
	Most popular $\alpha$ values are lies in between  $[0.01, 0.05]$ and taken in the account for our experiment too except one exception of $\alpha=0.001$. Let suppose if $\alpha=0.05$, expect to obtain sample means in the critical region $5\%$ of the time when the $H_0$ is true. Then, we can not determine that the null hypothesis is true, but if it falls in the critical region then it get reject. This is the reason, $\alpha$ value consider sometime as error rate. In our test result if by providing the $\epsilon$-value relaxation in $\alpha$, if the test is pass than result can accepted. The true nature of $\epsilon$-value relaxation is also need to explore. 

	\subsection{Kolmogorov-Smirnov test (KS-test) of generated datasets}
	Let suppose the sample of RN generator from our algorithm is $ r_1,r_2, .....,r_n $ where the $n$ represent the total number of sample of $RN$  generated. In KS-test, the sub-sample is being selected from the sample size in such a way that size of sub-sample $\le n$. The hypothesis suggests that if the subsample passes the test for some individual $\alpha$ value, it means the number passes the uniformity test, and the test can further also proceed for different sub-sample. The empirical $s_n(x)$ is depended upon the sample size taken for test and can define:

	\begin{equation}
	\label{val_eq1}
	s_n(s)=\left(\frac{|r_i,r_{i+1},...r_{i+j}| \le x}{N} \right) 
	where, \hspace{0.2 cm} N=\frac{n}{4} \hspace{0.2 cm} and \hspace{0.2 cm} N \subset n
	\end{equation}
	
	The KS-test define with the greatest absolute expectation among between $f(x)$ and $s_n(x)$, over the range of random variables and difference in parametric test (DPT) can define as
	
	\begin{equation}
	\label{val_eq2}
	DPT=|f(x)-s_n(x)| 
	\end{equation}
	
	(See Appendices A.2 for detail)

	\subsection{$\chi^{2}-test$ (See Appendices A.2 for detail)}
	
	Chi-Square goodness of fit test determines if a sample data have similarity with the  population. 	Once the $\chi^2$ value is calculated, then with respect to the degree of freedom and $\alpha$ value, the  individual $\chi^2$ value is being check and if the value is larger then the hypothesis is not reject.

	\subsection{Auto correlation test (See Appendices A.2 for detail)}
	
	It concerned with the dependencies between numbers in sequence to check whether it is any dependency on the number or not. 	Once the $Z_0$ value computed, it will compare with the Cumulative standardized normal distribution table and if $Z_0 \le Table(Z_0)$ w.r.t $\alpha$, then the conclusion can be made that there is no alpha and hypothesis should be accepted. 	Apart from the auto-correlation, if the $x$ and $y$  value is in same dimension (which is true in case of deployment), circular co-relation has been suggested for a periodic sequence in \cite{salivahanan2001digital}.

	\begin{figure}[t] 
		\centering
		\subfloat[ \label{1a}]{%
			\includegraphics[width=0.33\linewidth]{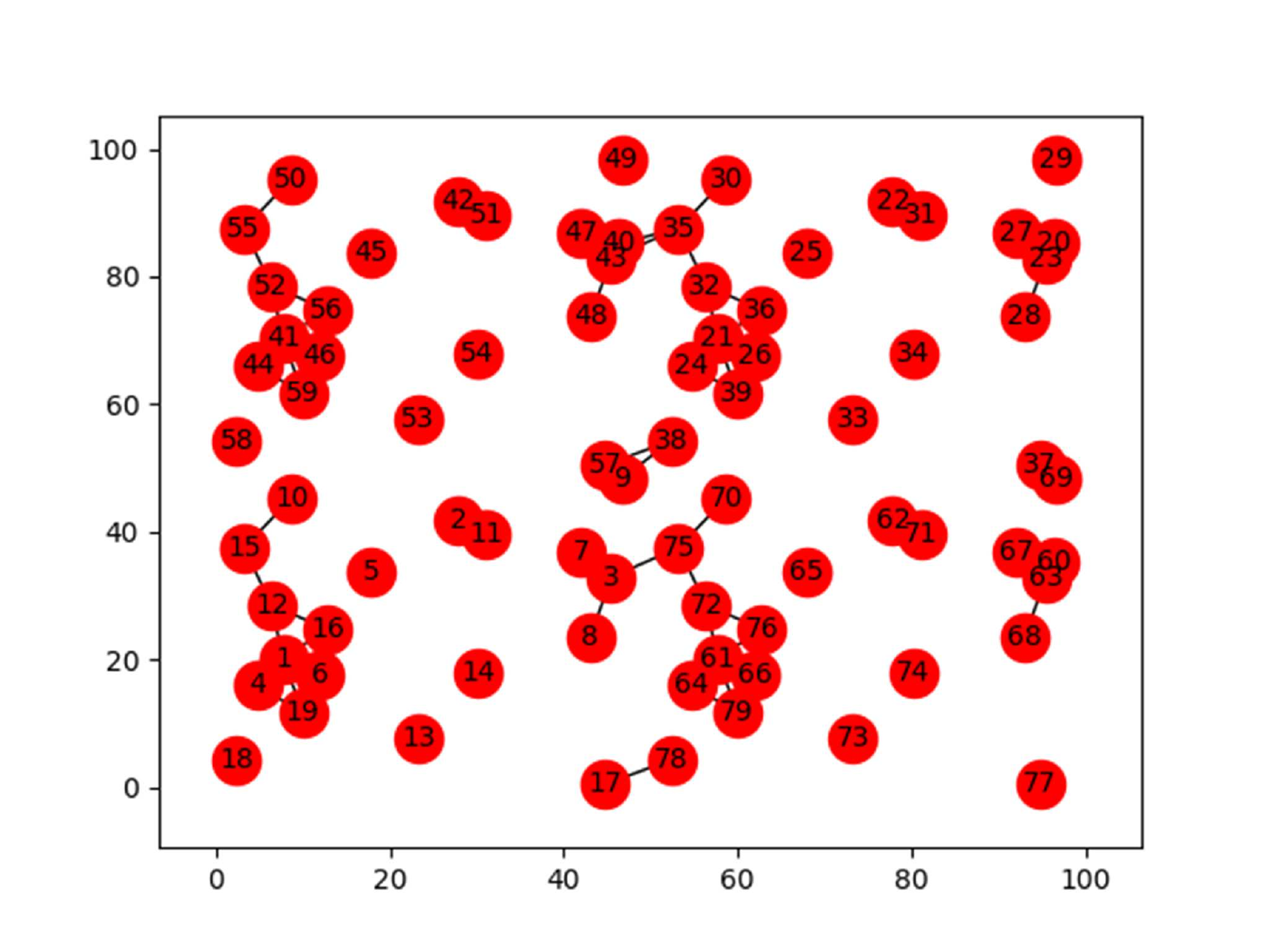}}
		\subfloat[ \label{1b}]{%
			\includegraphics[width=0.33\linewidth]{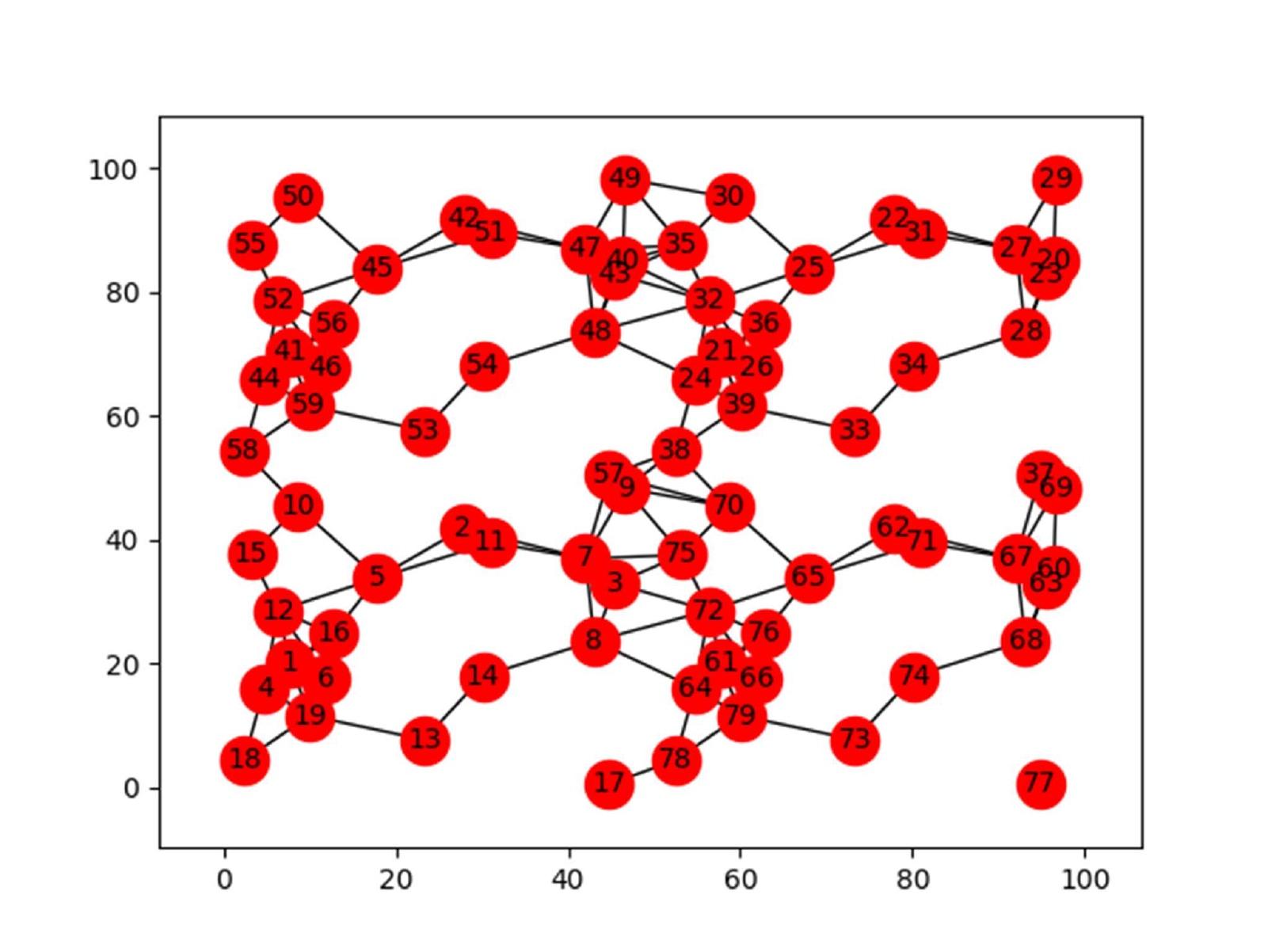}} 
		\subfloat[ \label{1c}]{%
			\includegraphics[width=0.33\linewidth]{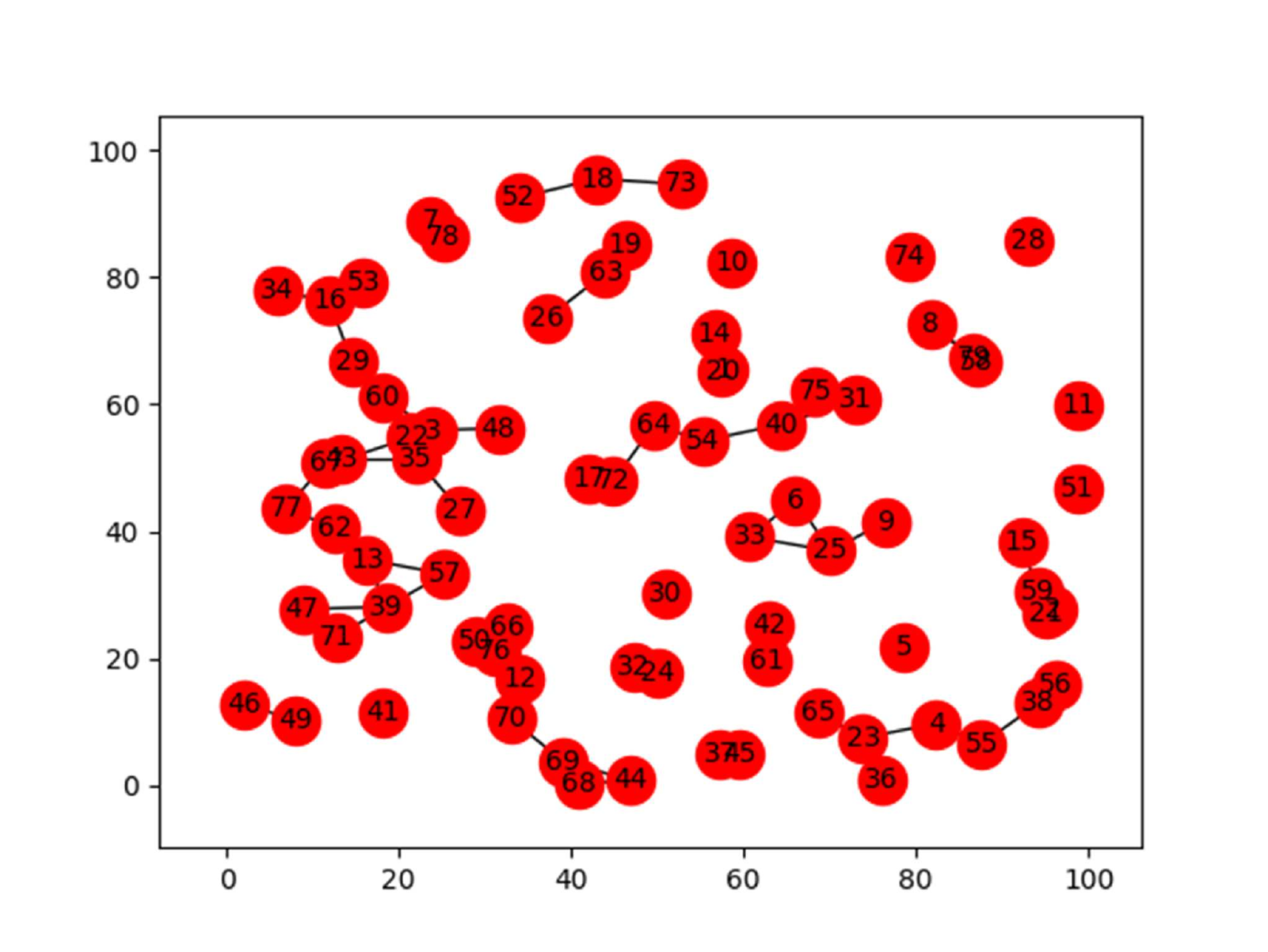}}
		\hfill
		\subfloat[ \label{1d}]{%
			\includegraphics[width=0.33\linewidth]{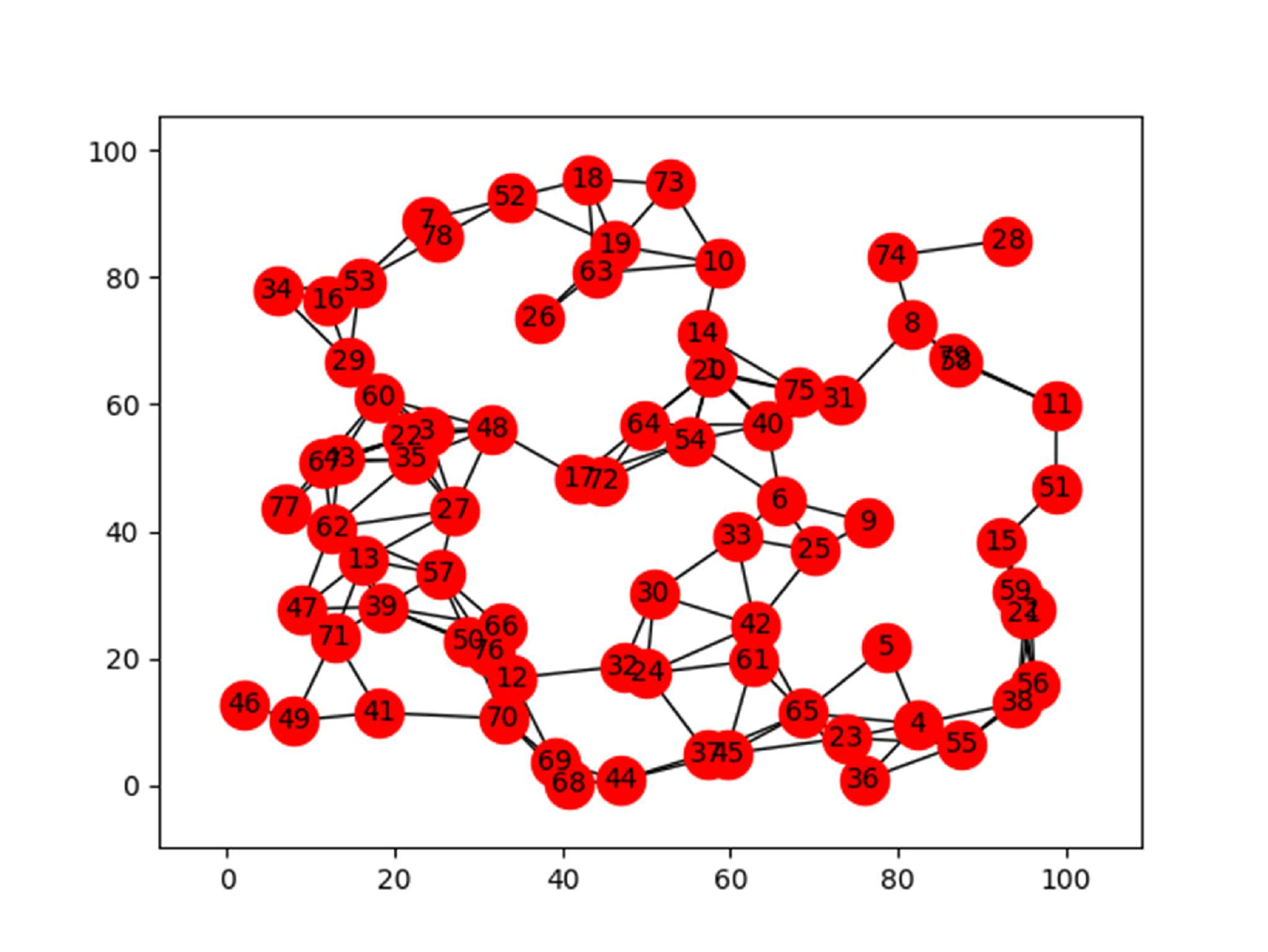}} 
		\subfloat[ \label{1e}]{%
			\includegraphics[width=0.33\linewidth]{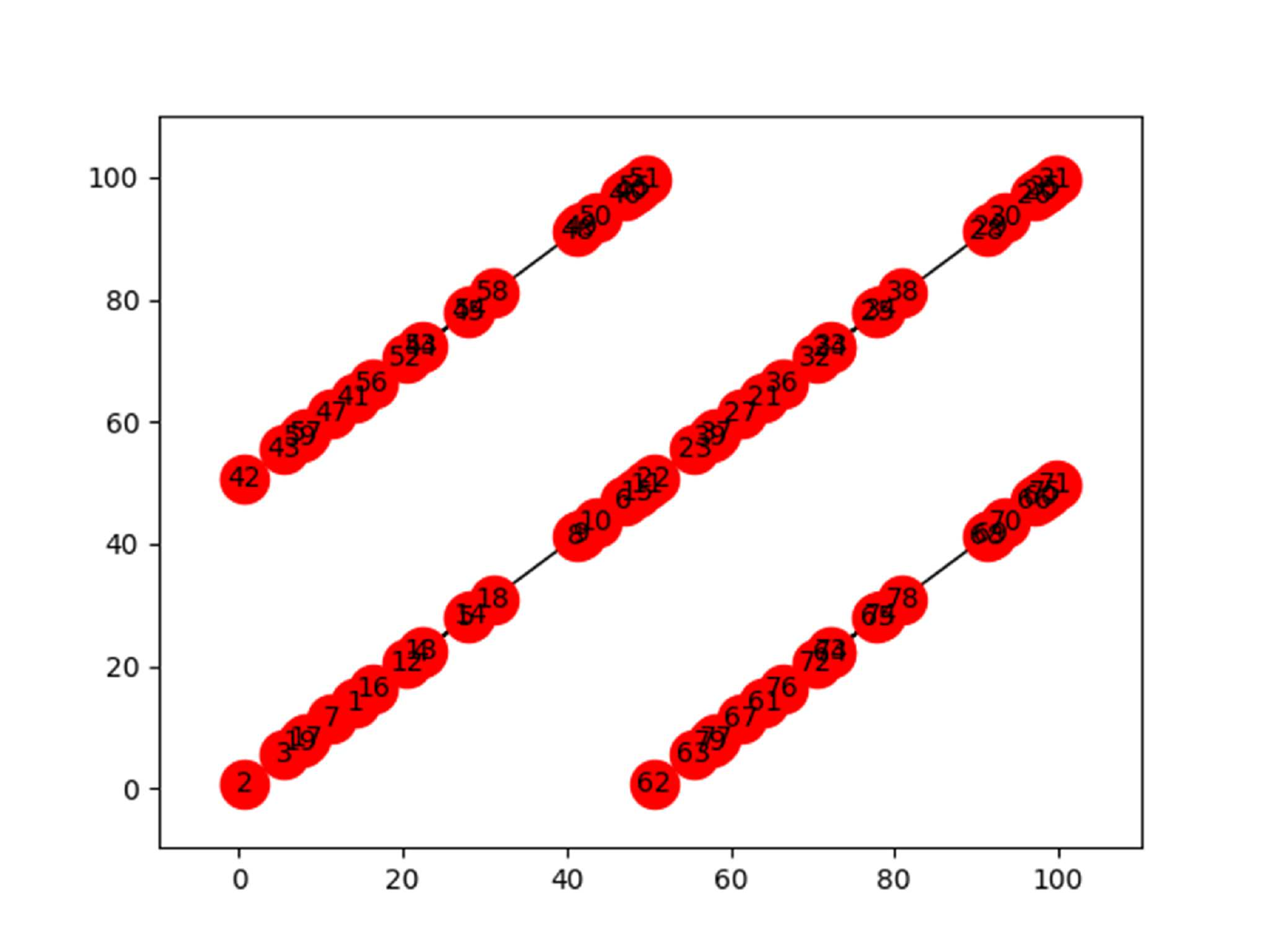}}
		\subfloat[ \label{1f}]{%
			\includegraphics[width=0.33\linewidth]{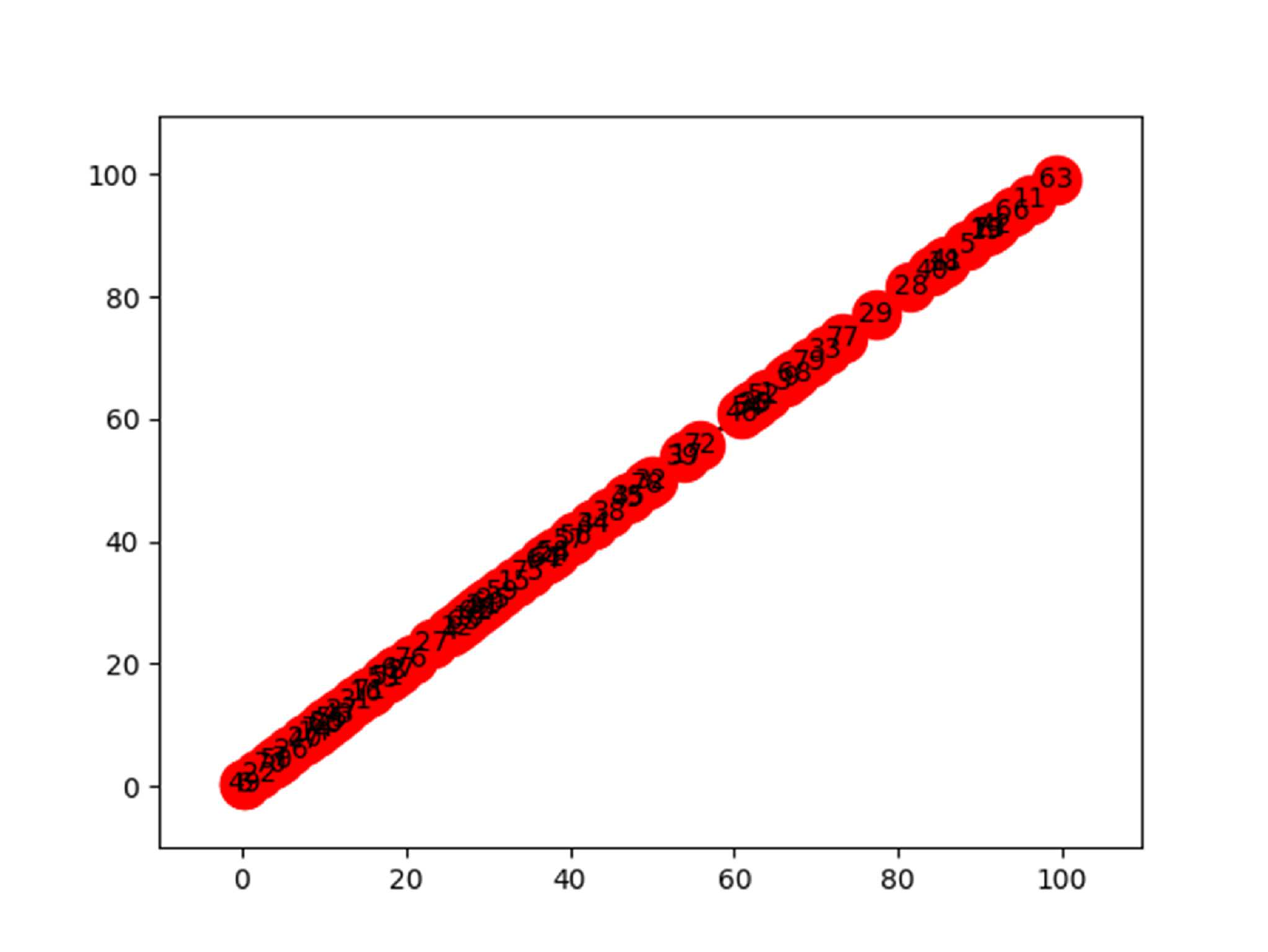}} 
		
		\caption{Deployment of the network in grid and non-grid modes with X[0]=43, a=3.359885666	b=1.902160583 (a) Transmission range=10 $(Grid-based)$ (b) Transmission range=15 $(Grid-based)$ (c) Transmission range= 10 $(Non Grid-based)$ (d) Transmission range=15 $(Non Grid-based)$ (e) a=c $(Grid-based)$ (f) a=c $(Non grid-based)$
		}
		\label{fig_2} 
	\end{figure}
	
	Test result of the sample data-set of experiment is available in Appendices A.3 Table.~\ref{table_1}.
	
	\subsection{Results}
\begin{figure}[t] 
		\centering
		\subfloat[ a.\label{2a}]{%
			\includegraphics[width=0.4\linewidth]{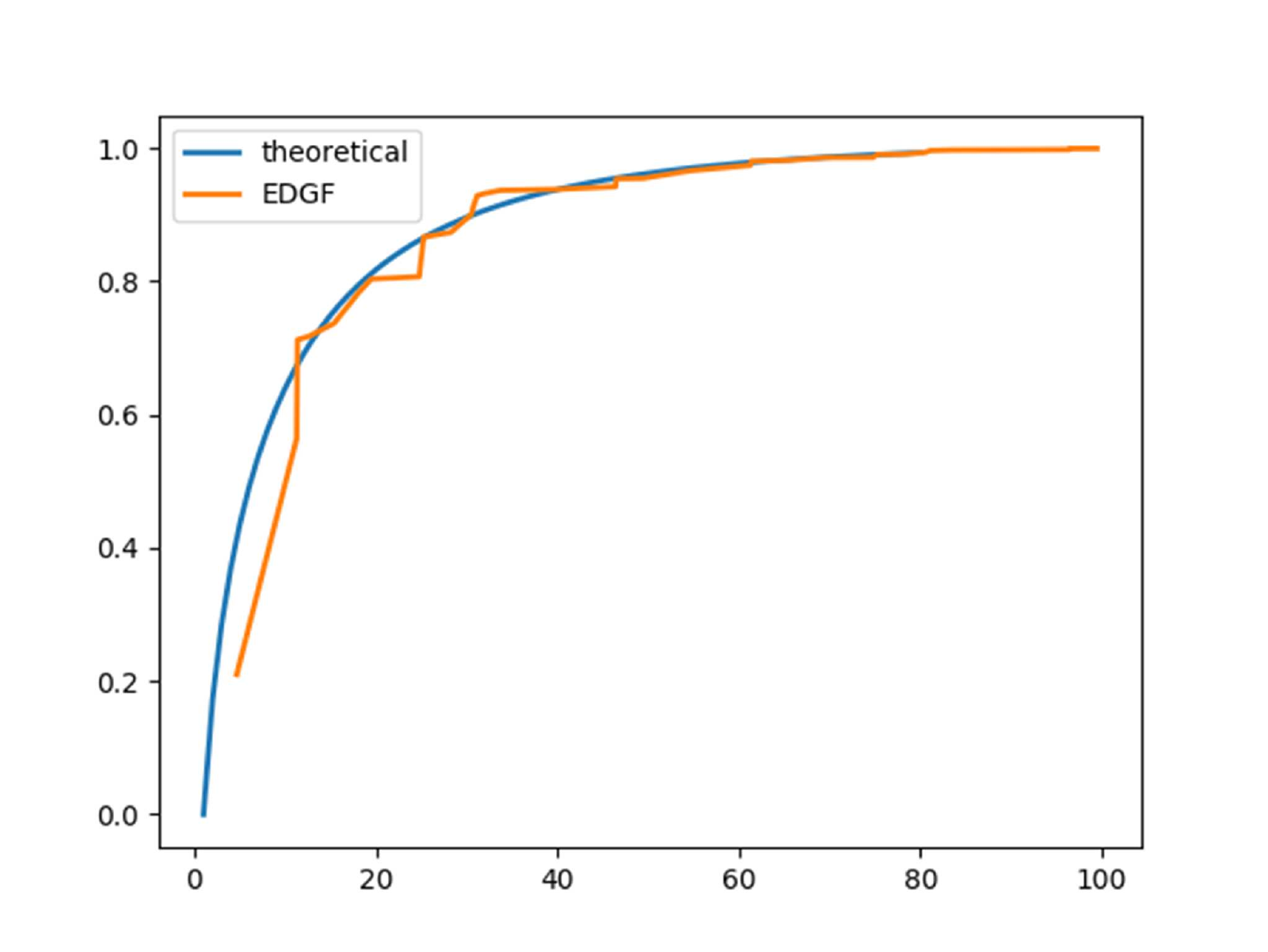}}
		\subfloat[ b.\label{2b}]{%
			\includegraphics[width=0.4\linewidth]{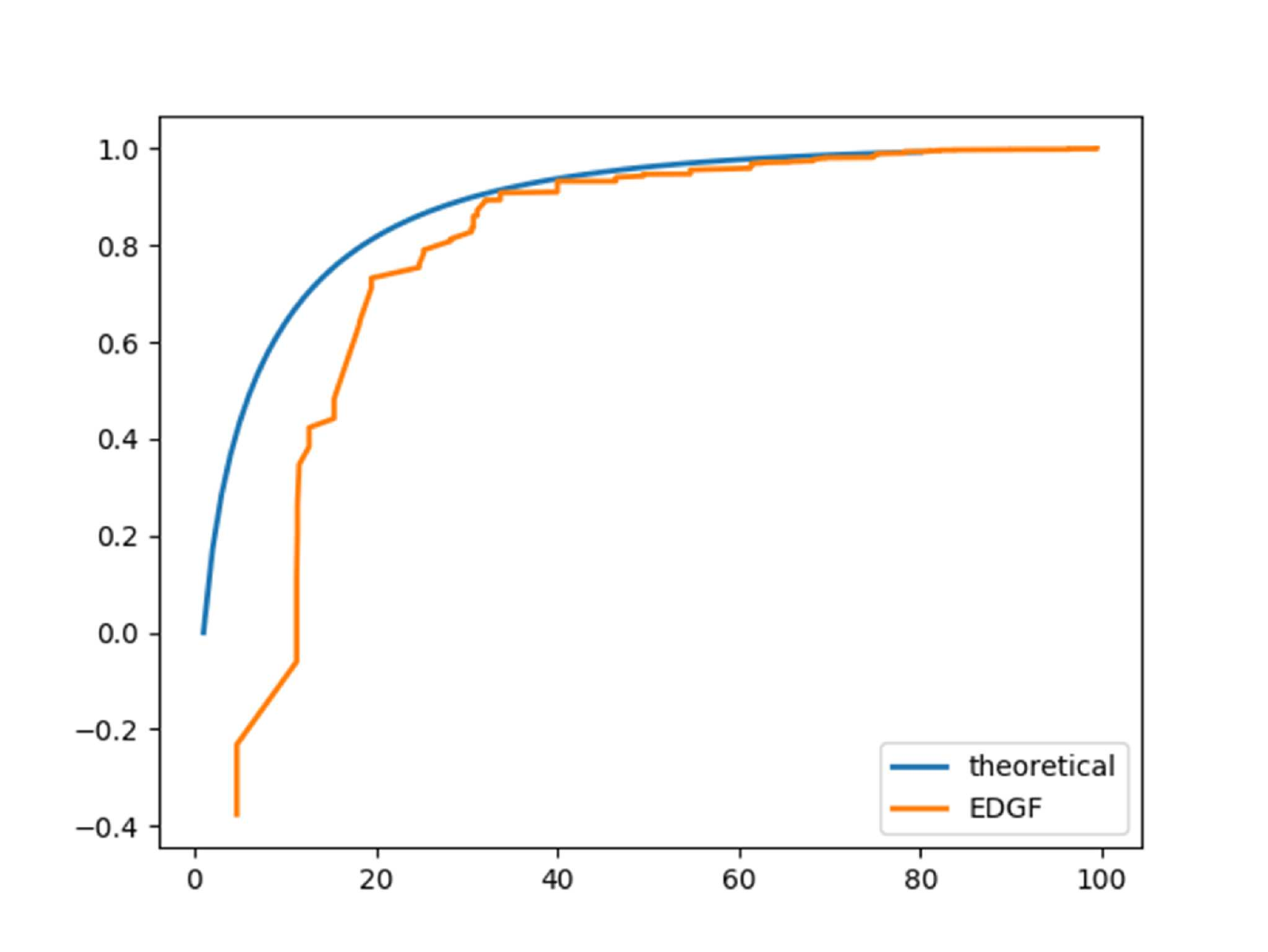}} 
		
		\subfloat[ c.\label{3b}]{%
			\includegraphics[width=0.4\linewidth]{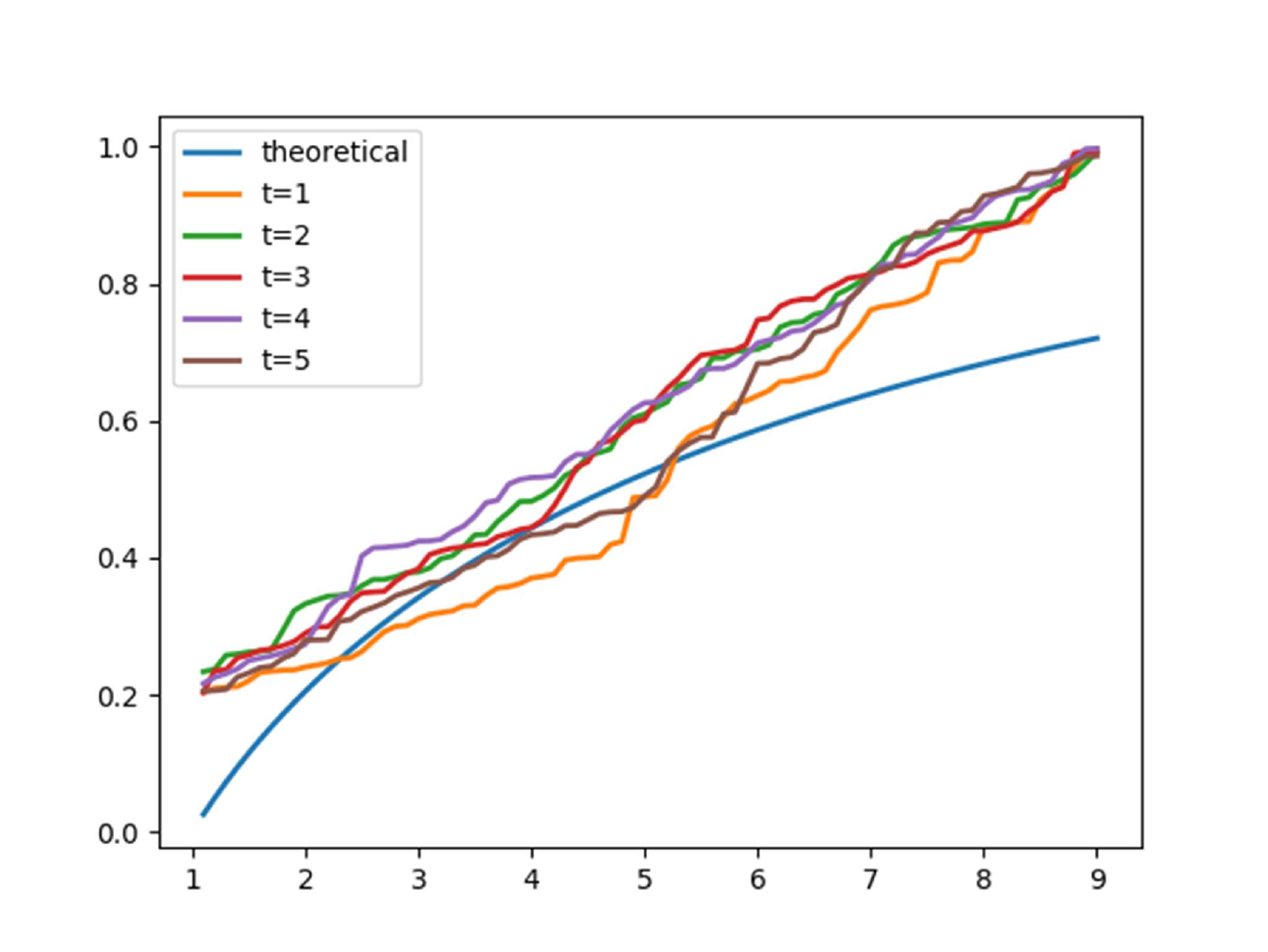}}
		\subfloat[ d. \label{3a}]{%
			\includegraphics[width=0.4\linewidth]{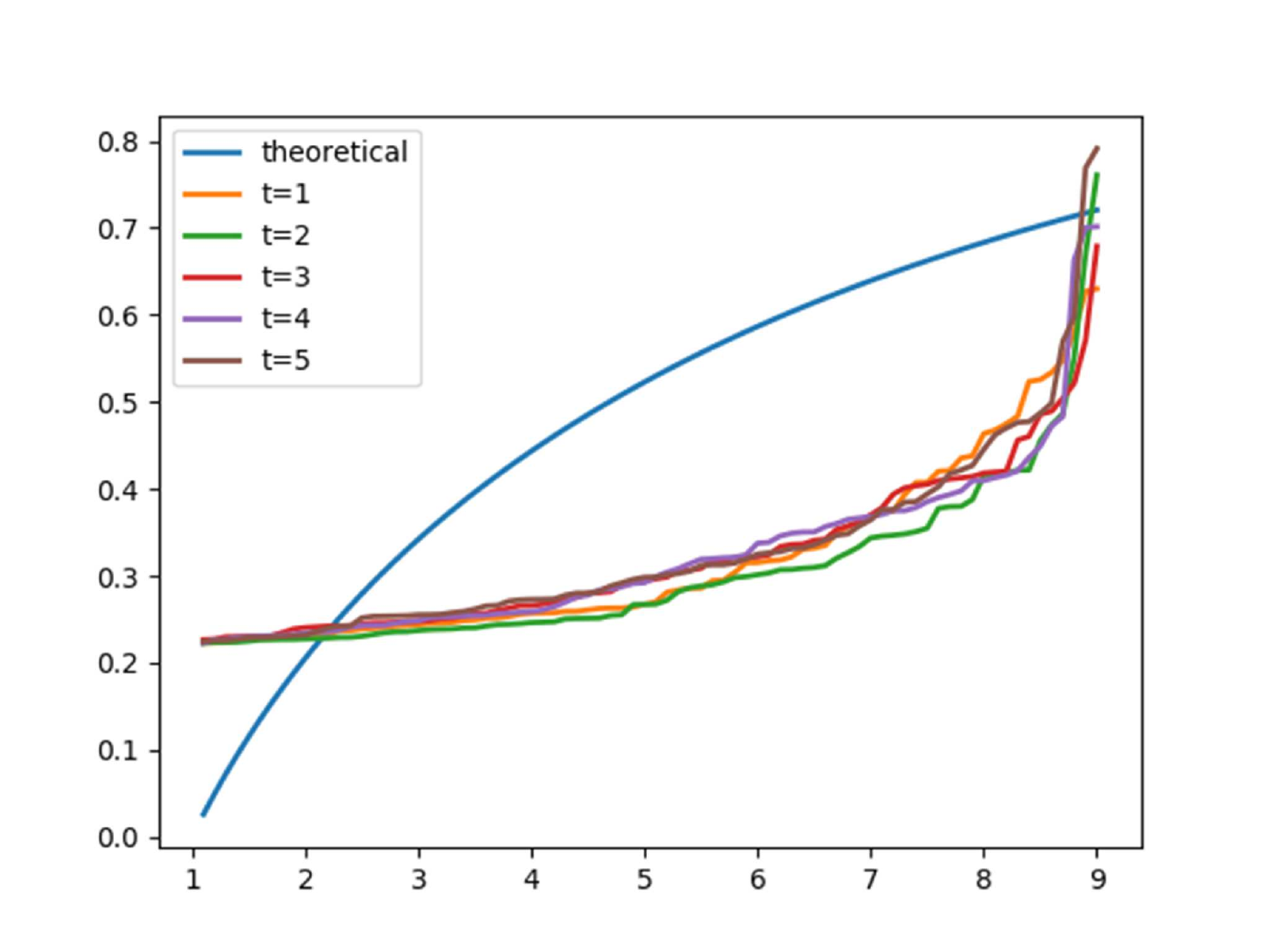}}
		
		\caption{Illustration of the Kolmogorov-Smirnov statistic for sensor deployment and packet generation  (a) Grid-based (b) Non-grid based  (c) Uniform distribution  (d)Exponential approach }
		\label{fig_3} 
	\end{figure}

	Fig. \ref{fig_2} represent both grid and non-grid based random deployment of sensor nodes in a 100 sq. m area with the seed value 43 and different transmission ranges. In Grid-based deployment, we generate random position in one quarter and repeat the same positions in remaining quarters. Grid-based deployment makes easier to test the data set, and satisfies the connectivity of the nodes and covering the area. Fig. \ref{1a} and \ref{1b} shows the grid-based deployment with transmission range 10 and 15 respectively. The observation here we found that the sensors are well connected when increases the transmission range.  The non-grid based random sensor deployment shown in Fig. \ref{2a} and \ref{2b} with transmission ranges 10 and 15. In non-grid based random deployment we found that the sensor placements are varied and the probability of placing more number of nodes in particular portion is high. The constant values such as $a$ and $c$ will get more impact to get the random deployment. We need to make some difference between these two constant values. If we make both $a$ and $c$ are same then the result will be like Fig. \ref{1e} and Fig.\ref{1f}, where Fig. \ref{1e} represents Grid-based deployment and Fig. \ref{1f} represents non-grid based deployment.
	
	In Fig.~\ref{fig_3}, K-S test statics of performed simulation have shown for 
	EDGF and cumulative distribution function (CDF) for deployment and packet generation data. The $DPT$ shown in equation.~\ref{val_eq2} can be easily observe in Fig.~\ref{fig_3}, as in Fig.~\ref{2a} and Fig.~\ref{2b}  the blue line shows the CDF based value of randomness and the orange line is able to replicate the uniformity in both grid and non-grid deployment. In Fig.~\ref{3a} and Fig.~\ref{3b}, the  depiction of packet generation for different time slots $t_1,t_2,..,t_5$ has shown. The packet data seems does not passing the uniformity in Fig.~\ref{3a} for some interval which shows the biasness of the of the function. Even though, for uniform distribution as in Fig.~\ref{3a}, reference CDF denoted with blue line have showing partial failure, the exponential packet generation value in Fig.~\ref{3b} is showing promising difference.

	\appendix
	\section{Appendices}
	\subsection{Nature of distribution of the $\lambda \rightarrow 0$ in exponential distribution}
	Let $X_1,X_2,..X_N$ is the independent exponentially distributed random variables with variable $\lambda _1, .., \lambda_{n}$. Then $min{X_1..,X_n}$
	is also exponentially distributed, r $\lambda= \lambda _1+..+\lambda_{n}$ 
	
	\begin{equation}
	P(min{X_1..X_n})=P({X_1>x..X_n}>x)= \Pi_{i=1}^{n}  P(X_i>x)
	\end{equation}
	
	\begin{equation}
	P(min{X_1..X_n})= \Pi_{i=1}^{n} exp \left(-x\sum_{i=1}^{n} \lambda_i \right)
	\end{equation}
	
	To ensure the distribution still obeying the exponential, the index of variable $k$ which nature $\lambda \rightarrow 0$ is define as  
	
	\begin{equation}
	P(k|X_k=min{X_1..X_n})=\frac{\lambda_k}{\lambda_1+\lambda_2+...+\lambda_n}
	\end{equation}

	\subsection{Validation Test Description}
	Let suppose $X$ is uniformly distributed over the unit interval of $[0,1]$ then cumulative distribution function (CDF) of $X$ can define as:  
	
	\begin{equation}
	\label{val_eq0}
	f(x)=\begin{cases}
	0: & \text{ $x<0$}\\
	x: & \text{$0\le x < 1$}\\
	1: & \text {$x \ge 1$}
	\end{cases}
	\end{equation}

	\subsubsection{Kolmogorov-Smirnov test:}
	
	Following steps require to follow:
	\begin{itemize}
		\item[step.1] Compute $D^+$
		\begin{equation}
		\label{step_eq1}
		D^+=\max\limits_{1\leq j\leq n} \Bigg\{ \frac {i} {n}-r_i \Bigg \}
		\end{equation}
		\begin{equation}
		\label{step_eq2}
		D^-=\max\limits_{1\leq j\leq n} \Bigg\{ r_i-\frac {i-1} {n} \Bigg \}
		\end{equation}

		\item[step.2] Compute max among $D^+$ and $D^-$ as 
		\begin{equation}
		\label{step_eq3}
		D= [ D^+,D^- ] 
		\end{equation}
		\item[step.3] Locate in KS-table critical value of $D_\alpha  $ for specified $\alpha$ in sample space. 
		
	\end{itemize}

	\subsubsection{$\chi^2$-test} 
	To perform the $\chi^2$ test, we need to perform following steps on the sample $RV$ for validation.
	\begin{itemize}
		\item[step.1] Compute $\chi^2$
		\begin{equation}
		\label{step_2_eq1}
		\chi^2= \sum_{i=1}^{n} \Bigg\{ \frac {(F_i-AE_i)^2} {AE_i} \Bigg \}
		\end{equation}
		where $F_i$ is frequency of number in the class $i^{th}$ class from uniform distribution. $AE_i$ is defined as absolute expected number in each class equal to $N/n$ for equally space class. 
		\item[step.2] Compute $\nu$
		
		\begin{equation}
		\label{step_2_eq2_2}
		\nu=number \hspace{0.2 cm} of \hspace{0.2 cm}i -1
		\end{equation}
		Sampling distribution of $\chi^2$ is approximately the $\chi^2$ distribution with $(n-1)$ degree of freedom.
		
		\subsubsection{autocorrelation and circular autocorrelation -test}
		
		To compute the autocorrelation between every $m$ number (m is known as lag), will start from the $i^{th}$ number with m as interval and can define as $i+m, i+2m, i+3m ...$. To proceeds with, the following step is being taken to compute the autocorrelation $\rho_{im}$ between numbers $R_i, R_{i+m}, R_{i+2m}, ....R_{i+(M+1)}$ is to be found where $M$ is largest integer as $i+(M+1)m\le N$.

		An non-zero auto-correlation implies lack of dependence following detailed test appropriate.
		
		$H:\rho_{im} = 0$ no correlation
		
		$H_I:\rho_{im}\neq 0$ correlation
		\begin{itemize}
			\item[step.1] Compute $Z_0$
			\begin{equation}
			\label{step_1_eq1_2}
			Z_0= \frac{\widehat{\rho}_{im}}{\delta_{ \widehat{\rho}_{im}}}
			\end{equation}
			
			\item[step.2] Compute $\widehat{\rho}_{im}$
			\begin{equation}
			\label{step_2_eq2}
			\widehat{\rho}_{im}= \frac{1}{M+1}\sum_{k=0}{M}\left[ R_{i+KM} R_{i+(K+1)m} \right]-0.25
			\end{equation}
			\item[step.3]
			\begin{equation}
			\label{step_3_eq2}
			{\delta_{ \widehat{\rho}_{im}}}= \sqrt{\frac{13M+7}{12(M+1)}}
			\end{equation}  
			
		\end{itemize}
		
		If two sequence of RN is periodic with same period $N$, then the circular correlation can define as 
		
		\begin{equation}
		\widehat{\rho}_{im}=\frac{1}{N} \sum_{k=0}^{N-1}{x(n)y(n-m)}
		\end{equation} 
		
		\begin{equation}
		{\delta_{ \widehat{\rho}_{im}}}= \sqrt{\frac{13N+7}{12(N+1)}}
		\end{equation} 
		
		\begin{equation*}
		where \hspace{0.2 cm} m=0,1,...,(N-1) \hspace{0.2 cm} 
		\end{equation*}
		\begin{equation*}
		Note:\text{Only $m$=0 considered in experiment with $\alpha=0.001$ for circulation coreelation}
		\end{equation*}

	\end{itemize}

	\subsection{Test case results Table.\ref{table_1}}
	\begin{sidewaystable}
		\tiny
		\centering
		\caption{Test case results}
		\label{table_1}
		\begin{tabular}{|l|l|l|l|l|l|l|l|l|l|l|l|l|l|l|}
			\hline
			\multicolumn{3}{|l|}{}                                                                                                  & \multicolumn{6}{c|}{\textbf{Isolated nodes in non-grid based deployment and Testing}}                                                                                                                                                                                                              & \multicolumn{6}{c|}{\textbf{Isolated nodes in grid based deployment and Testing}}                                                                                                                                                                                                                   \\ \hline
			\multicolumn{1}{|c|}{\textbf{X{[}0{]}}} & \multicolumn{1}{c|}{\textbf{a value}} & \multicolumn{1}{c|}{\textbf{c value}} & \multicolumn{1}{c|}{\textbf{TR=10}} & \multicolumn{1}{c|}{\textbf{TR=15}} & \multicolumn{1}{c|}{\textbf{TR=20}} & \multicolumn{1}{c|}{\textbf{KS-Test}} & \multicolumn{1}{c|}{\textbf{Chi2Test}} & \multicolumn{1}{c|}{\textbf{\begin{tabular}[c]{@{}c@{}}Auto\\ correlation\\ Test\end{tabular}}} & \multicolumn{1}{c|}{\textbf{TR=10}} & \multicolumn{1}{c|}{\textbf{TR=15}} & \multicolumn{1}{c|}{\textbf{TR=20}} & \multicolumn{1}{c|}{\textbf{KS-Test}} & \multicolumn{1}{c|}{\textbf{Chi2Test}} & \multicolumn{1}{c|}{\textbf{\begin{tabular}[c]{@{}c@{}}Auto-\\ correlation\\ Test\end{tabular}}} \\ \hline
			0                                       & 4.669202                              & 2.295587                              & 10                                  & 1                                   & 0                                   & Satisfied                             & Satisfied                              & Satisfied                                                                                       & 15                                  & 0                                   & 0                                   & Satisfied                             & Satisfied                              & Satisfied                                                                                        \\ \hline
			2                                       & 3.275823                              & 1.705211                              & 11                                  & 0                                   & 0                                   & Satisfied                             & Satisfied                              & Satisfied                                                                                       & 8                                   & 4                                   & 0                                   & Rejected                              & Satisfied                              & Satisfied                                                                                        \\ \hline
			3                                       & 2.80777                               & 1.324718                              & 8                                   & 2                                   & 0                                   & Rejected                              & Rejected                               & Satisfied                                                                                       & 13                                  & 2                                   & 0                                   & Satisfied                             & Satisfied                              & Satisfied                                                                                        \\ \hline
			5                                       & 2.584982                              & 3.141593                              & 11                                  & 1                                   & 1                                   & Satisfied                             & Satisfied                              & Satisfied                                                                                       & 12                                  & 0                                   & 0                                   & Satisfied                             & Satisfied                              & Satisfied                                                                                        \\ \hline
			7                                       & 2.295587                              & 4.669202                              & 1                                   & 0                                   & 0                                   & Satisfied                             & Satisfied                              & Satisfied                                                                                       & 11                                  & 0                                   & 0                                   & Satisfied                             & Satisfied                              & Satisfied                                                                                        \\ \hline
			12                                      & 3.141593                              & 2.584982                              & 5                                   & 2                                   & 1                                   & Rejected                              & Satisfied                              & Satisfied                                                                                       & 9                                   & 0                                   & 0                                   & Satisfied                             & Satisfied                              & Satisfied                                                                                        \\ \hline
			14                                      & 4.669202                              & 2.295587                              & 8                                   & 2                                   & 0                                   & Rejected                              & Rejected                               & Satisfied                                                                                       & 14                                  & 2                                   & 0                                   & Satisfied                             & Rejected                               & Satisfied                                                                                        \\ \hline
			24                                      & 1.324718                              & 2.80777                               & 11                                  & 2                                   & 2                                   & Satisfied                             & Satisfied                              & Satisfied                                                                                       & 11                                  & 0                                   & 0                                   & Satisfied                             & Rejected                               & Satisfied                                                                                        \\ \hline
			43                                      & 3.359886                              & 1.902161                              & 8                                   & 0                                   & 0                                   & Satisfied                             & Satisfied                              & Satisfied                                                                                       & 17                                  & 1                                   & 1                                   & Satisfied                             & Satisfied                              & Satisfied                                                                                        \\ \hline
			59                                      & 2.80777                               & 1.324718                              & 8                                   & 1                                   & 1                                   & Satisfied                             & Satisfied                              & Satisfied                                                                                       & 12                                  & 0                                   & 0                                   & Satisfied                             & Satisfied                              & Satisfied                                                                                        \\ \hline
			65                                      & 1.705211                              & 3.275823                              & 6                                   & 0                                   & 0                                   & Satisfied                             & Rejected                               & Satisfied                                                                                       & 8                                   & 0                                   & 0                                   & Satisfied                             & Satisfied                              & Satisfied                                                                                        \\ \hline
			70                                      & 4.669202                              & 2.295587                              & 11                                  & 3                                   & 0                                   & Satisfied                             & Satisfied                              & Satisfied                                                                                       & 7                                   & 0                                   & 0                                   & Rejected                              & Satisfied                              & Satisfied                                                                                        \\ \hline
			76                                      & 2.502908                              & 2.718282                              & 5                                   & 0                                   & 0                                   & Satisfied                             & Satisfied                              & Satisfied                                                                                       & 9                                   & 3                                   & 0                                   & Satisfied                             & Rejected                               & Satisfied                                                                                        \\ \hline
			87                                      & 2.80777                               & 1.324718                              & 9                                   & 2                                   & 0                                   & Rejected                              & Satisfied                              & Satisfied                                                                                       & 8                                   & 2                                   & 0                                   & Rejected                              & Satisfied                              & Satisfied                                                                                        \\ \hline
			144                                     & 2.685452                              & 1.618034                              & 8                                   & 1                                   & 0                                   & Satisfied                             & Rejected                               & Satisfied                                                                                       & 6                                   & 0                                   & 0                                   & Satisfied                             & Satisfied                              & Satisfied                                                                                        \\ \hline
			147                                     & 2.295587                              & 4.669202                              & 8                                   & 1                                   & 0                                   & Satisfied                             & Satisfied                              & Satisfied                                                                                       & 21                                  & 6                                   & 0                                   & Satisfied                             & Rejected                               & Satisfied                                                                                        \\ \hline
			192                                     & 1.324718                              & 2.80777                               & 13                                  & 3                                   & 1                                   & Satisfied                             & Satisfied                              & Satisfied                                                                                       & 8                                   & 0                                   & 0                                   & Satisfied                             & Satisfied                              & Satisfied                                                                                        \\ \hline
			251                                     & 2.718282                              & 2.502908                              & 9                                   & 1                                   & 0                                   & Rejected                              & Satisfied                              & Satisfied                                                                                       & 4                                   & 0                                   & 0                                   & Rejected                              & Satisfied                              & Satisfied                                                                                        \\ \hline
			365                                     & 3.359886                              & 1.902161                              & 4                                   & 1                                   & 0                                   & Rejected                              & Rejected                               & Satisfied                                                                                       & 16                                  & 1                                   & 0                                   & Satisfied                             & Satisfied                              & Satisfied                                                                                        \\ \hline
			1111                                    & 2.584982                              & 3.141593                              & 9                                   & 3                                   & 0                                   & Satisfied                             & Satisfied                              & Satisfied                                                                                       & 8                                   & 0                                   & 0                                   & Satisfied                             & Satisfied                              & Satisfied                                                                                        \\ \hline
		\end{tabular}
	\end{sidewaystable}
	
	\subsection{Packet generation data for exponential distribution Table.\ref{table_2}}
	
	\begin{table}
		\tiny
		\caption{Exponential packet generation data}
		\label{table_2}
		\begin{longtable}{|c|l|l|l|l|l|l|l|l|l|l|}
			
			\hline
			\textbf{Node ID} & \multicolumn{5}{l|}{\textbf{Exponential distribution}}                   & \multicolumn{5}{l|}{\textbf{Uniform distribution}}                       \\ \hline
			\textbf{}        & \textbf{t=1} & \textbf{t=2} & \textbf{t=3} & \textbf{t=4} & \textbf{t=5} & \textbf{t=1} & \textbf{t=2} & \textbf{t=3} & \textbf{t=4} & \textbf{t=5} \\ \hline
			\textbf{1}       & 3.63         & 2.93         & 3.41         & 2.59         & 2.26         & 6.06         & 7.55         & 4.44         & 2.26         & 3.1          \\ \hline
			\textbf{2}       & 2.37         & 2.88         & 3.18         & 2.27         & 2.42         & 5.86         & 6.91         & 2.36         & 3.43         & 6.93         \\ \hline
			\textbf{3}       & 3.18         & 2.28         & 2.45         & 3.38         & 2.54         & 2.41         & 3.59         & 7.47         & 4.18         & 9.41         \\ \hline
			\textbf{4}       & 4.84         & 2.29         & 2.52         & 4.21         & 4.18         & 2.54         & 4.03         & 8.9          & 8.86         & 8.74         \\ \hline
			\textbf{5}       & 4.07         & 3.8          & 3.22         & 2.33         & 2.66         & 8.35         & 7.04         & 2.78         & 4.81         & 3.45         \\ \hline
			\textbf{6}       & 2.42         & 3.2          & 2.31         & 2.57         & 7.69         & 7            & 2.64         & 4.35         & 9.97         & 4.35         \\ \hline
			\textbf{7}       & 2.57         & 7.61         & 2.57         & 7            & 2.55         & 9.96         & 4.34         & 9.93         & 4.24         & 9.61         \\ \hline
			\textbf{8}       & 5.23         & 2.38         & 2.94         & 3.46         & 2.66         & 3.17         & 6.1          & 7.67         & 4.84         & 3.56         \\ \hline
			\textbf{9}       & 2.44         & 3.34         & 2.49         & 3.79         & 3.19         & 7.37         & 3.85         & 8.32         & 6.97         & 2.53         \\ \hline
			\textbf{10}      & 2.29         & 2.51         & 4.11         & 3.9          & 3.41         & 3.99         & 8.79         & 8.5          & 7.56         & 4.47         \\ \hline
			\textbf{11}      & 2.59         & 2.27         & 2.41         & 3.14         & 7.91         & 2.34         & 3.39         & 6.8          & 9.97         & 4.37         \\ \hline
			\textbf{12}      & 2.58         & 2.23         & 2.27         & 2.43         & 3.25         & 2.04         & 2.38         & 3.49         & 7.13         & 3.07         \\ \hline
			\textbf{13}      & 2.37         & 2.86         & 3.09         & 4.83         & 2.29         & 5.78         & 6.63         & 9.41         & 2.54         & 4.01         \\ \hline
			\textbf{14}      & 2.51         & 4.17         & 4.05         & 3.75         & 3.13         & 8.86         & 8.72         & 8.26         & 6.76         & 9.86         \\ \hline
			\textbf{15}      & 6.27         & 2.51         & 4.14         & 3.98         & 3.57         & 4.01         & 8.83         & 8.62         & 7.93         & 5.67         \\ \hline
			\textbf{16}      & 2.84         & 2.99         & 3.78         & 3.19         & 2.29         & 6.29         & 8.32         & 6.96         & 2.5          & 3.9          \\ \hline
			\textbf{17}      & 2.49         & 3.88         & 3.37         & 2.53         & 4.45         & 8.47         & 7.45         & 4.1          & 9.14         & 9.65         \\ \hline
			\textbf{18}      & 5.34         & 2.4          & 3.05         & 4.36         & 4.76         & 3.3          & 6.52         & 9.06         & 9.37         & 2.4          \\ \hline
			\textbf{19}      & 2.27         & 2.44         & 3.33         & 2.48         & 3.76         & 3.56         & 7.37         & 3.84         & 8.28         & 6.84         \\ \hline
			\textbf{20}      & 3.15         & 2.24         & 2.31         & 2.58         & 2.24         & 2.12         & 2.66         & 4.41         & 2.17         & 2.8          \\ \hline
			\textbf{21}      & 2.33         & 2.67         & 2.46         & 3.5          & 2.73         & 4.88         & 3.68         & 7.77         & 5.17         & 4.65         \\ \hline
			\textbf{22}      & 2.62         & 2.35         & 2.75         & 2.7          & 2.54         & 2.92         & 5.29         & 5.02         & 4.15         & 9.31         \\ \hline
			\textbf{23}      & 4.68         & 2.25         & 2.35         & 2.76         & 2.72         & 2.21         & 2.94         & 5.32         & 5.14         & 4.56         \\ \hline
			\textbf{24}      & 2.61         & 2.31         & 2.57         & 6.63         & 2.54         & 2.63         & 4.33         & 9.9          & 4.14         & 9.28         \\ \hline
			\textbf{25}      & 4.63         & 2.24         & 2.3          & 2.56         & 5.97         & 2.11         & 2.63         & 4.31         & 9.81         & 3.85         \\ \hline
			\textbf{26}      & 2.49         & 3.78         & 3.18         & 2.27         & 2.42         & 8.31         & 6.92         & 2.37         & 3.46         & 7.04         \\ \hline
			\textbf{27}      & 3.22         & 2.33         & 2.66         & 2.43         & 3.27         & 2.78         & 4.82         & 3.5          & 7.18         & 3.22         \\ \hline
			\textbf{28}      & 2.39         & 2.98         & 3.7          & 3.04         & 4.22         & 6.25         & 8.17         & 6.47         & 8.91         & 8.89         \\ \hline
			\textbf{29}      & 4.2          & 4.15         & 4.01         & 3.65         & 2.96         & 8.84         & 8.66         & 8.08         & 6.17         & 7.9          \\ \hline
			\textbf{30}      & 3.56         & 2.82         & 2.93         & 3.39         & 2.56         & 5.6          & 6.04         & 7.5          & 4.27         & 9.69         \\ \hline
			\textbf{31}      & 5.48         & 2.42         & 3.21         & 2.31         & 2.59         & 3.45         & 7.01         & 2.67         & 4.46         & 2.32         \\ \hline
			\textbf{32}      & 2.26         & 2.4          & 3.07         & 4.49         & 5.7          & 3.31         & 6.55         & 9.17         & 9.75         & 3.65         \\ \hline
			\textbf{33}      & 2.45         & 3.46         & 2.66         & 2.43         & 3.28         & 7.67         & 4.83         & 3.51         & 7.22         & 3.34         \\ \hline
			\textbf{34}      & 2.41         & 3.1          & 5.04         & 2.34         & 2.73         & 6.66         & 9.52         & 2.9          & 5.2          & 4.74         \\ \hline
			\textbf{35}      & 2.64         & 2.39         & 2.99         & 3.75         & 3.13         & 3.23         & 6.27         & 8.26         & 6.76         & 9.86         \\ \hline
			\textbf{36}      & 6.3          & 2.51         & 4.18         & 4.09         & 3.85         & 4.02         & 8.87         & 8.77         & 8.42         & 7.29         \\ \hline
			\textbf{37}      & 3.3          & 2.44         & 3.36         & 2.52         & 4.27         & 3.58         & 7.43         & 4.05         & 8.97         & 9.07         \\ \hline
			\textbf{38}      & 4.38         & 4.86         & 2.3          & 2.55         & 4.99         & 9.43         & 2.6          & 4.21         & 9.5          & 2.81         \\ \hline
			\textbf{39}      & 2.33         & 2.68         & 2.48         & 3.66         & 2.98         & 4.91         & 3.79         & 8.11         & 6.26         & 8.22         \\ \hline
			\textbf{40}      & 3.73         & 3.09         & 4.86         & 2.3          & 2.54         & 6.63         & 9.43         & 2.58         & 4.17         & 9.36         \\ \hline
			\textbf{41}      & 4.75         & 2.27         & 2.42         & 3.21         & 2.31         & 2.36         & 3.45         & 7.01         & 2.68         & 4.47         \\ \hline
			\textbf{42}      & 2.59         & 2.27         & 2.42         & 3.2          & 2.3          & 2.36         & 3.44         & 6.98         & 2.57         & 4.13         \\ \hline
			\textbf{43}      & 2.53         & 4.55         & 6.79         & 2.54         & 4.87         & 9.22         & 9.92         & 4.19         & 9.43         & 2.61         \\ \hline
			\textbf{44}      & 2.3          & 2.55         & 5.21         & 2.38         & 2.92         & 4.24         & 9.6          & 3.15         & 6.01         & 7.41         \\ \hline
			\textbf{45}      & 3.35         & 2.51         & 4.04         & 3.71         & 3.05         & 3.96         & 8.69         & 8.18         & 6.52         & 9.05         \\ \hline
			
		\end{longtable}
	\end{table}
	\begin{table}
		\tiny
		\begin{longtable}{|c|l|l|l|l|l|l|l|l|l|l|}
			\hline
			\textbf{Node ID} & \multicolumn{5}{l|}{\textbf{Exponential distribution}}                   & \multicolumn{5}{l|}{\textbf{Uniform distribution}}                       \\ \hline
			\textbf{}        & \textbf{t=1} & \textbf{t=2} & \textbf{t=3} & \textbf{t=4} & \textbf{t=5} & \textbf{t=1} & \textbf{t=2} & \textbf{t=3} & \textbf{t=4} & \textbf{t=5} \\ \hline
			\textbf{46}      & 4.35         & 4.74         & 2.27         & 2.41         & 3.12         & 9.35         & 2.34         & 3.37         & 6.74         & 9.78         \\ \hline
			\textbf{47}      & 5.8          & 2.47         & 3.57         & 2.84         & 2.99         & 3.73         & 7.93         & 5.67         & 6.27         & 8.25         \\ \hline
			\textbf{48}      & 3.74         & 3.12         & 5.71         & 2.46         & 3.47         & 6.73         & 9.75         & 3.66         & 7.69         & 4.9          \\ \hline
			\textbf{49}      & 2.67         & 2.47         & 3.62         & 2.91         & 3.31         & 3.76         & 8.02         & 5.98         & 7.31         & 3.64         \\ \hline
			\textbf{50}      & 2.45         & 3.44         & 2.63         & 2.36         & 2.8          & 7.62         & 4.67         & 2.99         & 5.51         & 5.76         \\ \hline
			\textbf{51}      & 2.86         & 3.07         & 4.56         & 7.02         & 2.55         & 6.57         & 9.23         & 9.93         & 4.25         & 9.61         \\ \hline
			\textbf{52}      & 5.26         & 2.39         & 2.97         & 3.6          & 2.88         & 3.2          & 6.19         & 7.98         & 5.86         & 6.91         \\ \hline
			\textbf{53}      & 3.17         & 2.26         & 2.41         & 3.08         & 4.77         & 2.33         & 3.33         & 6.62         & 9.38         & 2.42         \\ \hline
			\textbf{54}      & 2.28         & 2.45         & 3.42         & 2.61         & 2.3          & 3.63         & 7.58         & 4.55         & 2.62         & 4.27         \\ \hline
			\textbf{55}      & 2.56         & 5.51         & 2.43         & 3.24         & 2.36         & 9.7          & 3.48         & 7.11         & 3.01         & 5.56         \\ \hline
			\textbf{56}      & 2.81         & 2.9          & 3.24         & 2.36         & 2.8          & 5.92         & 7.11         & 2.99         & 5.51         & 5.76         \\ \hline
			\textbf{57}      & 2.86         & 3.07         & 4.6          & 2.23         & 2.27         & 6.58         & 9.26         & 2.04         & 2.38         & 3.5          \\ \hline
			\textbf{58}      & 2.43         & 3.27         & 2.39         & 2.99         & 3.76         & 7.18         & 3.23         & 6.28         & 8.28         & 6.84         \\ \hline
			\textbf{59}      & 3.15         & 2.24         & 2.3          & 2.54         & 4.7          & 2.1          & 2.58         & 4.16         & 9.33         & 2.26         \\ \hline
			\textbf{60}      & 2.26         & 2.37         & 2.89         & 3.21         & 2.32         & 3.11         & 5.9          & 7.03         & 2.74         & 4.67         \\ \hline
			\textbf{61}      & 2.63         & 2.36         & 2.82         & 2.92         & 3.36         & 3.02         & 5.59         & 6.02         & 7.43         & 4.03         \\ \hline
			\textbf{62}      & 2.52         & 4.21         & 4.19         & 4.13         & 3.94         & 8.91         & 8.88         & 8.81         & 8.57         & 7.76         \\ \hline
			\textbf{63}      & 3.5          & 2.72         & 2.6          & 2.29         & 2.52         & 5.14         & 4.53         & 2.54         & 4.03         & 8.9          \\ \hline
			\textbf{64}      & 4.21         & 4.18         & 4.1          & 3.85         & 3.32         & 8.87         & 8.77         & 8.43         & 7.33         & 3.71         \\ \hline
			\textbf{65}      & 2.46         & 3.55         & 2.8          & 2.85         & 3.03         & 7.87         & 5.5          & 5.71         & 6.42         & 8.74         \\ \hline
			\textbf{66}      & 4.07         & 3.8          & 3.21         & 2.32         & 2.62         & 8.34         & 7.03         & 2.72         & 4.61         & 2.8          \\ \hline
			\textbf{67}      & 2.33         & 2.67         & 2.47         & 3.57         & 2.82         & 4.89         & 3.73         & 7.91         & 5.62         & 6.1          \\ \hline
			\textbf{68}      & 2.94         & 3.47         & 2.67         & 2.47         & 3.64         & 7.69         & 4.91         & 3.77         & 8.07         & 6.13         \\ \hline
			\textbf{69}      & 2.95         & 3.51         & 2.74         & 2.65         & 2.4          & 7.78         & 5.21         & 4.76         & 3.29         & 6.47         \\ \hline
			\textbf{70}      & 3.04         & 4.22         & 4.2          & 4.16         & 4.02         & 8.91         & 8.9          & 8.84         & 8.68         & 8.14         \\ \hline
			\textbf{71}      & 3.68         & 3.01         & 3.94         & 3.49         & 2.71         & 6.37         & 8.56         & 7.75         & 5.08         & 4.34         \\ \hline
			\textbf{72}      & 2.57         & 6.69         & 2.54         & 4.72         & 2.26         & 9.91         & 4.16         & 9.34         & 2.31         & 3.28         \\ \hline
			\textbf{73}      & 2.4          & 3.03         & 4.13         & 3.93         & 3.48         & 6.44         & 8.81         & 8.56         & 7.73         & 5.03         \\ \hline
			\textbf{74}      & 2.7          & 2.54         & 4.89         & 2.31         & 2.58         & 4.19         & 9.45         & 2.65         & 4.38         & 2.05         \\ \hline
			\textbf{75}      & 2.23         & 2.28         & 2.46         & 3.5          & 2.73         & 2.44         & 3.68         & 7.78         & 5.18         & 4.67         \\ \hline
			\textbf{76}      & 2.63         & 2.36         & 2.81         & 2.88         & 3.15         & 3            & 5.53         & 5.84         & 6.83         & 2.06         \\ \hline
			\textbf{77}      & 2.23         & 2.28         & 2.48         & 3.68         & 3.01         & 2.47         & 3.8          & 8.14         & 6.36         & 8.55         \\ \hline
			\textbf{78}      & 3.93         & 3.48         & 2.7          & 2.53         & 4.62         & 7.73         & 5.02         & 4.14         & 9.27         & 2.08         \\ \hline
			\textbf{79}      & 2.23         & 2.29         & 2.51         & 4.1          & 3.85         & 2.53         & 3.99         & 8.77         & 8.43         & 7.33         \\ \hline
			\textbf{80}      & 3.32         & 2.46         & 3.53         & 2.78         & 2.78         & 3.71         & 7.84         & 5.4          & 5.4          & 5.4          \\ \hline
		\end{longtable}
	\end{table}


\bibliographystyle{elsarticle-num}
	\bibliography{dinesh} 
\end{document}